\documentclass[runningheads]{llncs}
\usepackage[T1]{fontenc}
\usepackage{booktabs}
\usepackage{graphicx}
\usepackage{bibunits}
\usepackage{multirow}
\usepackage[misc]{ifsym}
\usepackage{booktabs}
\usepackage{amsmath,amssymb,amsfonts}
\usepackage[dvipsnames]{xcolor}
\usepackage{bm}
\newcommand{\ie}{\textit{i.\,e.}}

\usepackage[colorlinks=true,linkcolor=blue,citecolor=blue,urlcolor=blue]{hyperref}
\usepackage{color}

\definecolor{mygreen}{RGB}{87, 176, 98}
\definecolor{myred}{RGB}{235, 70, 56}
\definecolor{myyellow}{RGB}{245, 198, 79}
\begin{document}
\title{
TransNuSeg: A Lightweight Multi-Task Transformer for Nuclei Segmentation
}
\titlerunning{TransNuSeg}
\author{
Zhenqi He\inst{1}\and Mathias Unberath\inst{2}\and Jing Ke\inst{3}\and Yiqing Shen\inst{2}\textsuperscript{(\Letter)}
}
\authorrunning{Z.He et al.}
\institute{
\textsuperscript{1}The University of Hong Kong\\
\textsuperscript{2}Johns Hopkins University\\
\textsuperscript{3}Shanghai Jiao Tong University\\
\email{yshen92@jhu.edu}
}
\maketitle              

\begin{abstract}
%
Nuclei appear small in size, yet, in real clinical practice, the global spatial information and correlation of the color or brightness contrast between nuclei and background, have been considered a crucial component for accurate nuclei segmentation.
However, the field of automatic nuclei segmentation is dominated by Convolutional Neural Networks (CNNs), meanwhile, the potential of the recently prevalent Transformers has not been fully explored, which is powerful in capturing local-global correlations.  
To this end, we make the first attempt at a pure Transformer framework for nuclei segmentation, called \texttt{TransNuSeg}.
Different from prior work, we decouple the challenging nuclei segmentation task into an intrinsic multi-task learning task, where a tri-decoder structure is employed for nuclei instance, nuclei edge, and clustered edge segmentation respectively. 
To eliminate the divergent predictions from different branches in previous work, a novel self distillation loss is introduced to explicitly impose consistency regulation between branches.
Moreover, to formulate the high correlation between branches and also reduce the number of parameters, an efficient attention sharing scheme is proposed by partially sharing the self-attention heads amongst the tri-decoders.
Finally, a token MLP bottleneck replaces the over-parameterized Transformer bottleneck for a further reduction in model complexity.
Experiments on two datasets of different modalities, including MoNuSeg have shown that our methods can outperform state-of-the-art counterparts such as CA\textsuperscript{2.5}-Net by 2-3\% Dice with 30\% fewer parameters. 
In conclusion, \texttt{TransNuSeg} confirms the strength of Transformer in the context of nuclei segmentation, which thus can serve as an efficient solution for real clinical practice. 
Code is available at \url{https://github.com/zhenqi-he/transnuseg}.

\end{abstract}

\section{Introduction}
Accurate cancer diagnosis, grading, and treatment decisions from medical images heavily rely on the analysis of underlying complex nuclei structures~\cite{method_for_nuclei}. 
Yet, due to the numerous nuclei contained in a digitized whole-slide image (WSI), or even in an image patch of deep learning input, dense annotation of nuclei contouring is extremely time-consuming and labor-expensive~\cite{a_review_and_comparison_of_breast_tumor}.
Consequently, automated nuclei segmentation approaches have emerged to satisfy a broad range of computer-aided diagnostic systems, where the deep learning methods,  particularly the convolutional neural networks~\cite{U-Net,Unet++,NucleiSegNet,HistoSeg,maskGANet} have received notable attention due to their simplicity and generalization ability.

\begin{figure}[t!]
\centering
\includegraphics[width=0.9\textwidth]{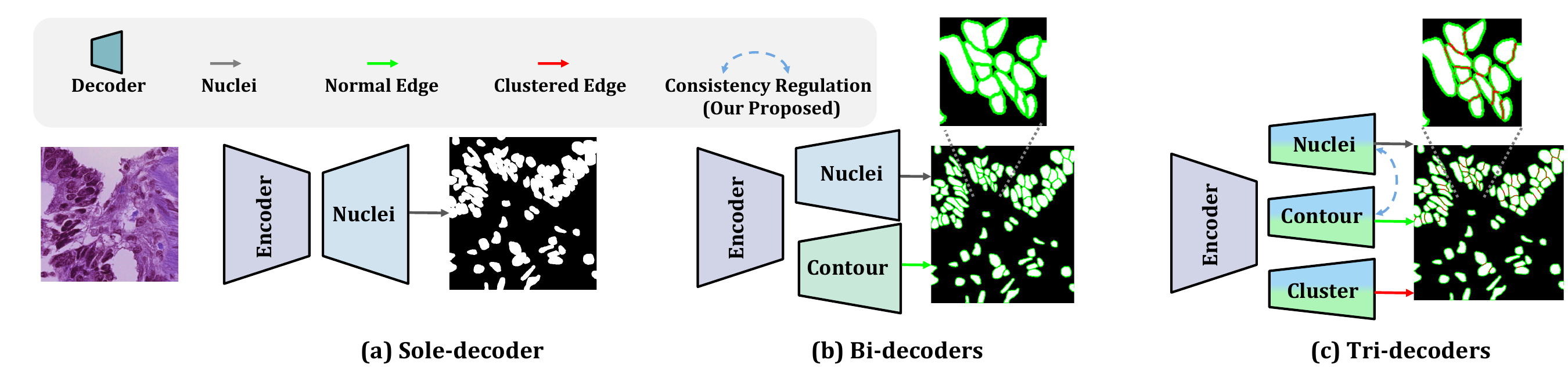}
\caption{Semantic illustrations of the nuclei segmentation networks with different numbers of decoders. 
(a) Sole-decoder to perform a single task of nuclei segmentation. 
(b) Bi-decoder to segment nuclei and locate nuclei edges simultaneously. 
(c) Tri-decoder with the third encoder path to specify the challenging clustered edge (ours), where the consistency regularization is designed across the predictions from the other two branches (dashed line).
} 
\label{fig:intro}
\end{figure}

In the literature work, the sole-decoder design in these UNet variants (Fig.~\ref{fig:intro} (a)) is susceptible to failures in splitting densely clustered nuclei when precise edge information is absent. 
Hence, deep contour-aware neural network (DCAN)~\cite{DCAN} with bi-decoder structure achieves improved instance segmentation performance by adopting multi-task learning,  in which one decoder learns to segment the nuclei and the other recognizes edges as described in Fig.~\ref{fig:intro} (b). 
Similarly, CIA-Net~\cite{CIA-net} extends DCAN with an extra information aggregator to fuse the features from two decoders for more precise segmentation. 
Much recently, CA\textsuperscript{2.5}-Net~\cite{CA2.5net} shows identifying the clustered edges in a multiple-task learning manner can achieve higher performance, and thereby proposes an extra output path to learn the segmentation of clustered edges explicitly.
%
%
A significant drawback of the aforementioned multi-decoder networks is the ignorance of the prediction consistency between branches, resulting in sub-optimal performance and missing correlations between the learned branches.
Specifically, a prediction mismatch between the nuclei and edge branches is observed in previous work~\cite{ClusterSeg}, implying a direction for performance improvement.
To narrow this gap, we propose a consistency distillation between the branches, as shown by the dashed line in Fig.~\ref{fig:intro} (c).
Furthermore, to resolve the cost of involving more decoders, we propose an attention sharing scheme, along with an efficient token MLP bottleneck~\cite{token-MLP}, which can both reduce the number of parameters.

Additionally, existing methods are CNN-based, and their intrinsic convolution operation fails to capture global spatial information or the correlation amongst nuclei~\cite{DA-Net}, which domain experts rely heavily on for accurate nuclei allocation. 
It suggests the presence of long-range correlation in practical nuclei segmentation tasks.
Inspired by the capability in long-range global context capturing by Transformers~\cite{Attention_is_all_u_need}, we make the first attempt to construct a tri-decoder based Transformer model to segment nuclei. 
In short, our major contributions are three-fold:
(1) 
We propose a novel multi-task framework for nuclei segmentation, namely \texttt{TransNuSeg}, as the first attempt at a fully Swin-Transformer driven architecture for nuclei segmentation. 
(2) 
To alleviate the prediction inconsistency between branches, we propose a novel self distillation loss that regulates the consistency between the nuclei decoder and normal edge decoder. 
(3) 
We propose an innovative attention sharing scheme that shares attention heads amongst all decoders.
By leveraging the high correlation between tasks, it can communicate the learned features efficiently across decoders and sharply reduce the number of parameters.
Furthermore, the incorporation of a light-weighted MLP bottleneck leads to a sharp reduction of parameters at no cost of performance decline.

\section{Methodology}

\begin{figure}[b!]
\centering
\includegraphics[width=0.8\textwidth]{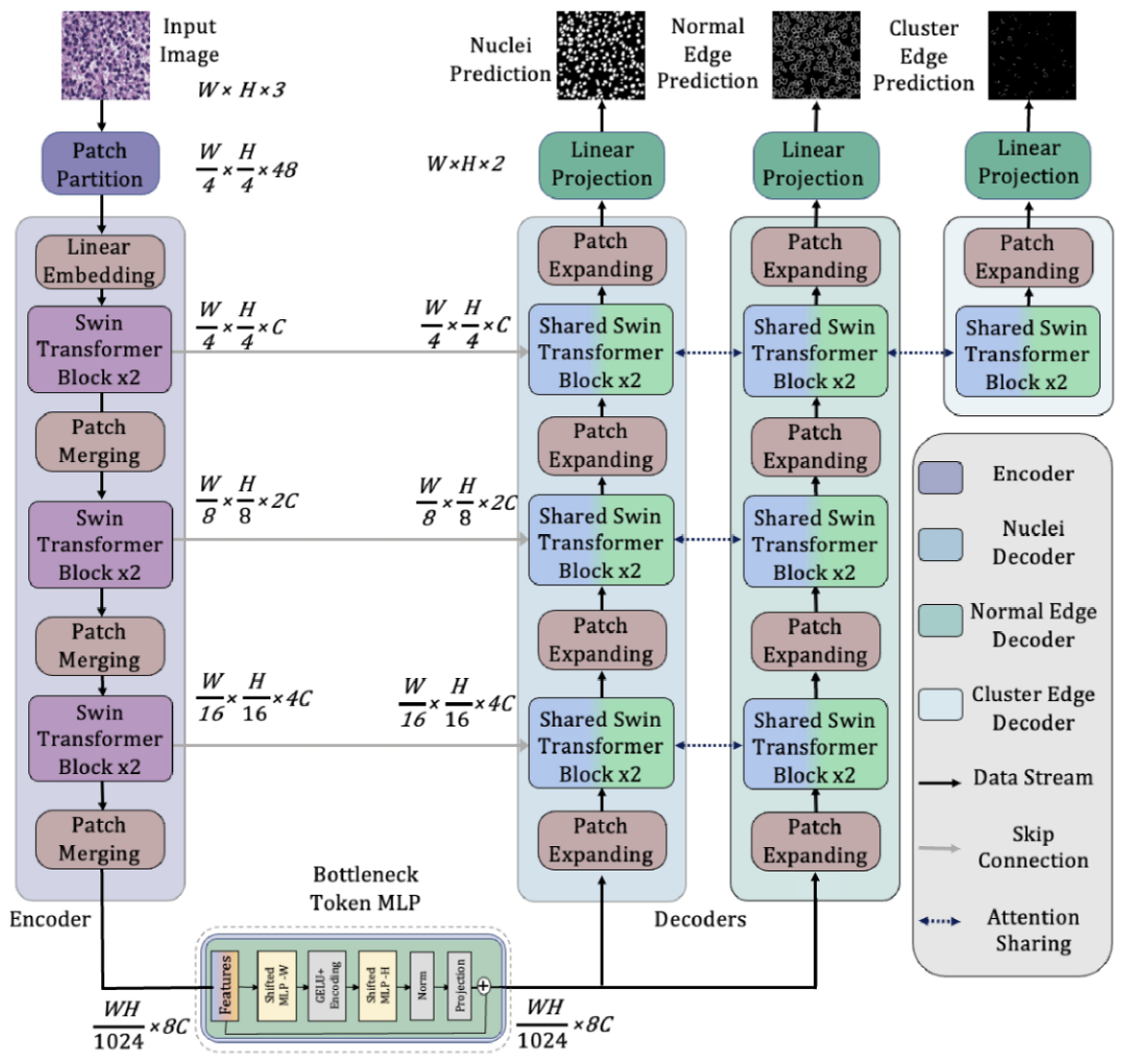}
\caption{
The overall framework of the proposed \texttt{TransNuSeg} of three output branches to separate the nuclei, normal edges, and cluster edges, respectively. 
In the novel design, a pre-defined proportion of the attention heads are shared between the decoders via the proposed sharing scheme, which considerably reduces the number of parameters and enables more efficient information communication. 
}\label{fig:arc}
\end{figure}

\subsubsection{Network Architecture Overview.} 
Fig.~\ref{fig:arc} illustrates the overall architecture of the proposed multi-task tri-decoder Transformer network, named \texttt{TransNuSeg}.
Both the encoder and decoders utilize the Swin Transformer~\cite{Swin-Transformer} as the building blocks to capture the long-range feature correlations in the nuclei segmentation context. 
Our network consists of three individual output decoder paths for nuclei segmentation, normal edges segmentation, and clustered edges segmentation.
Given the high dependency between edge and clustered edge, we are inspired to propose a novel attention sharing scheme, which can communicate the information and share learned features across decoders while also reducing the number of parameters.
Additionally, a token MLP bottleneck is incorporated to further increase the model efficiency.

\subsubsection{Attention Sharing Scheme.}
\begin{figure}[b!]
\centering
\includegraphics[width=0.8\textwidth]{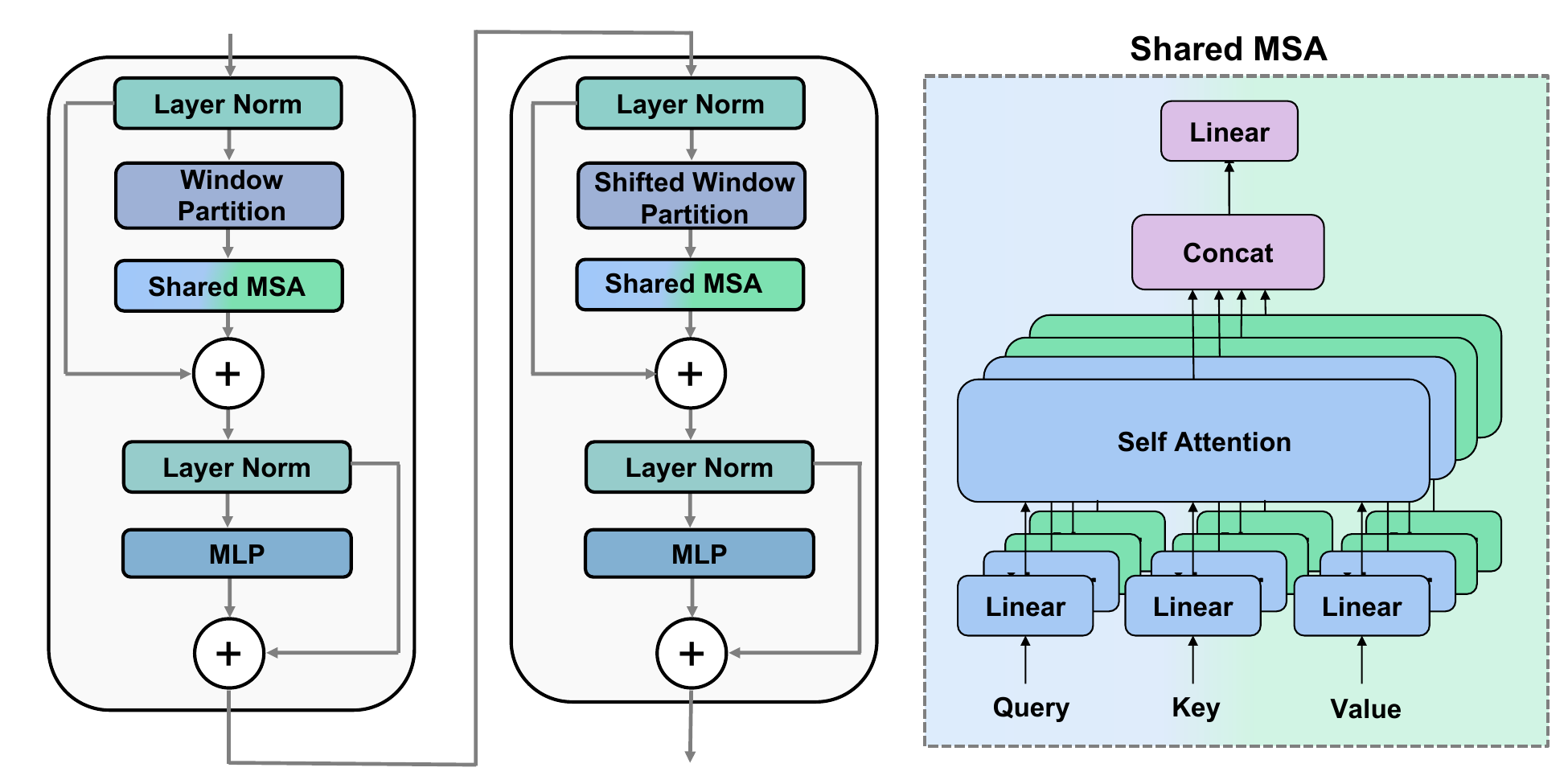}
\caption{A schematic illustration of the proposed Attention Sharing scheme.}\label{fig:attn}
\end{figure}

To capture the strong correlation between nuclei segmentation and contour segmentation between multiple decoders~\cite{shen2022federated}, we introduce a novel attention sharing scheme that is designed as an enhancement to the multi-headed self-attention (MSA) module in the plain Transformer~\cite{Attention_is_all_u_need}.
Based on the attention sharing scheme, we design a shared MSA module, which is similar in structure to vanilla MSA.
Specifically, it consists of a \texttt{LayerNorm} layer~\cite{ba2016layer}, residual connection, and feed-forward layer.
Innovatively, it differs from the vanilla MSA by sharing a proportion of globally-shared self-attention (SA) heads amongst all the parallel Transformer blocks in decoders, while keeping the remaining SA heads unshared \ie~learn the weights separately. 
A schematic illustration of the shared MSA module in the Swin Transformer block is demonstrated in Fig.~\ref{fig:attn}, as is formally formulated as follows:  
\begin{equation}
    \text{Shared-MSA}(\mathbf{z})=
    \Big[
    \text{SA}_1^s(\mathbf{z}),\cdots,\text{SA}_m^s(\mathbf{z}),
    \text{SA}_1^u(\mathbf{z}),\cdots,\text{SA}_n^u(\mathbf{z})
    \Big] \mathbf{U}_{\text{MSA}}^u
    ,
\end{equation}
$[\cdot]$ writes for the concatenation, $\text{SA}(\cdot)$ denotes the self-attention head whose output dimension is $D_h$, and $\mathbf{U}_{\text{MSA}}^u\in \mathbb{R}^{(m+n)\cdot D_h \times D}$ is a learnable matrix.
The superscript $s$ and $u$ refer to the globally-shared and unshared weights across all decoders, respectively.

\subsubsection{Token MLP Bottleneck.}
To reduce the complexity of the model, we leverage a token MLP bottleneck as a light-weight alternative for the Swin Transformer bottleneck.
Specifically, this approach involves shifting the latent features extracted by the encoder via two MLP blocks across the width and height channels, respectively~\cite{token-MLP}.
The objective of this process is to attend to specific areas, which mimics the shifted window attention mechanism in Swin Transformer~\cite{Swin-Transformer}.
The shifted features are then projected by a learnable MLP and normalized through a \texttt{LayerNorm}~\cite{ba2016layer} before being fed to a reprojection MLP layer. 

\subsubsection{Consistency Self Distillation.}
To alleviate the inconsistency between the contour generated from the nuclei segmentation prediction and the predicted edge, we propose a novel consistency self distillation loss, denoted as $\mathcal{L}_{SD}$.
Formally, this regularization is defined as the dice loss between the contour generated from the nuclei branch prediction ($y_{n}$) using the Sobel operation ($\texttt{sobel}(y_{n})$) and the predicted edges $y_{e}$ from the normal edge decoder. 
Specifically, the self distillation loss $\mathcal{L}_{D}$ is formulated by $\mathcal{L}_{sd} = \texttt{Dice}(\texttt{sobel}(y_{n}),y_{e})$.

\subsubsection{Multi-Task Learning Objective.}
We employ a multi-task learning paradigm to train the tri-decoder network, aiming to improve model performance by leveraging the additional supervision signal from edges.
Particularly, the nuclei semantic segmentation is considered the primary task, while the normal edge and clustered edge semantic segmentation are viewed as auxiliary tasks. 
All decoder branches follow a uniform scheme that combines the cross-entropy loss and the dice loss, with the balancing coefficients set to $0.60$ and $0.40$ respectively, as previous work \cite{CA2.5net}. 
Subsequently, the overall loss $\mathcal{L}$ is calculated as a weighted summation of semantic nuclei mask loss ($\mathcal{L}_{n}$), normal edge loss ($\mathcal{L}_{e}$), and clustered edge loss ($\mathcal{L}_{c}$), and the self distillation loss ($\mathcal{L}_{SD}$) \ie

$\mathcal{L} = \gamma_{n}\cdot\mathcal{L}_{n} + \gamma_{e}\cdot\mathcal{L}_{e} + \gamma_{c}\cdot\mathcal{L}_{c} + \gamma_{sd}\cdot\mathcal{L}_{sd}$,
where coefficients $\gamma_{n}$, $\gamma_{e}$ and $\gamma_{c}$ are set to $0.30$, $0.35$, $0.35$ respectively, and $\gamma_{sd}$ is initially set to $1$ with a $0.3$ decrease for every 10 epochs until it reaches $0.4$.

\begin{table}[ht!]
\centering
\caption{
Quantitative comparisons with counterparts. The best performance with respect to each metric is highlighted in \textbf{boldface}. 
}
\label{tab:result}
\begin{tabular}{l|l|ccccc}
\toprule
Dataset & Methods & DSC (\%) & F1 (\%) & Acc (\%) & IoU (\%) & ErCnt (\%)\\ 
\hline
\multirow{6}{*}{Microscopy}& UNet & $85.51$\tiny{$\pm0.35$} & $91.05$\tiny{$\pm0.13$} & $92.19$\tiny{$\pm0.20$} & $85.44$\tiny{$\pm0.29$} & $55.2$\tiny{$\pm2.7$}\\
& UNet++ & $94.14$\tiny{$\pm0.58$}  & $92.34$\tiny{$\pm0.63$} & $93.87$\tiny{$\pm0.61$} & $86.20$\tiny{$\pm1.02$} & $69.3$\tiny{$\pm1.4$}\\ 
& TransUNet & $94.14$\tiny{$\pm0.47$} & $92.31$\tiny{$\pm0.34$} & $93.76$\tiny{$\pm0.50$} & $86.16$\tiny{$\pm0.56$} & $51.9$\tiny{$\pm1.0$}\\
& SwinUNet & $96.05$\tiny{$\pm0.27$} & $95.02$\tiny{$\pm0.23$} & $96.08$\tiny{$\pm0.23$} & $91.06$\tiny{$\pm0.43$} &$31.2$\tiny{$\pm0.6$} \\ 
& CA\textsuperscript{2.5}-Net & $91.08$\tiny{$\pm0.49$} & $90.05$\tiny{$\pm0.27$} & $93.40$\tiny{$\pm0.14$} & $86.89$\tiny{$\pm0.87$} & $18.6$\tiny{$\pm1.3$}\\ 
\cline{2-7}
& \textbf{Ours} & \bm{$97.01$}\tiny{$\pm0.74$} & \bm{$96.67$}\tiny{$\pm0.60$} & \bm{$97.11$}\tiny{$\pm1.02$} & \bm{$92.97$}\tiny{$\pm0.41$} & \bm{$9.78$}\tiny{$\pm2.1$}\\ 
\hline
\multirow{6}{*}{Histology} & UNet & $80.97$\tiny{$\pm0.75$} & $72.17$\tiny{$\pm0.49$} & $90.14$\tiny{$\pm0.24$} & $61.63$\tiny{$\pm0.36$} & $45.7$\tiny{$\pm1.6$}\\ 
& UNet++ & $87.10$\tiny{$\pm0.16$} & $75.20$\tiny{$\pm0.19$} & $91.34$\tiny{$\pm0.14$}& $62.89$\tiny{$\pm0.27$} & $38.0$\tiny{$\pm2.4$}\\  
& TransUNet & $85.80$\tiny{$\pm0.20$} & $72.87$\tiny{$\pm0.49$}  & $90.53$\tiny{$\pm0.27$} & $60.21$\tiny{$\pm0.46$} & $35.2$\tiny{$\pm0.8$}\\ 
& SwinUNet & $88.73$\tiny{$\pm0.90$} & $78.11$\tiny{$\pm1.88$} & $91.23$\tiny{$\pm0.73$} & $64.41$\tiny{$\pm0.15$} & $27.6$\tiny{$\pm2.3$}\\ 
& CA\textsuperscript{2.5}-Net & $86.74$\tiny{$\pm0.18$} & $77.42$\tiny{$\pm0.30$} & $91.52$\tiny{$\pm0.78$} & $66.79$\tiny{$\pm0.34$} & $23.7$\tiny{$\pm0.7$}\\ 
\cline{2-7} 
& \textbf{Ours} & \bm{$90.81$}\tiny{$\pm0.22$} & \bm{$81.52$}\tiny{$\pm0.44$} & \bm{$92.77$}\tiny{$\pm0.64$} & \bm{$69.49$}\tiny{$\pm0.17$} & \bm{$11.4$}\tiny{$\pm1.1$}\\ 
\bottomrule
\end{tabular}
\end{table}

\begin{table}[b!]
\centering
\caption{
Comparison of the model complexity in terms of the number of parameters, FLOPs, as well as the training cost in the form of the averaged training time per epoch.
The average training time is computed using the same batch size for both datasets, with the first number indicating the averaged time on the Fluorescence Microscopy Image Dataset and the second on the Histology Image Dataset. 
The token MLP bottleneck and attention sharing scheme are denoted as `MLP', and `AS', respectively.
}
\label{tab:cost}
\begin{tabular}{l|c|c|c}
\toprule
Methods & \#Params~($\times10^6$) & FLOPs~($\times10^9$) & Training~(s) \\ 
\hline
UNet \cite{U-Net} & $31.04$ & $219.03$ & $43.4$/$27.7$ \\
UNet++ \cite{Unet++} & $9.05$ & $135.72$ & $41.8$/$31.7$ \\ 
TransUNet \cite{Trans-Unet} & $67.87$ & $129.97$ & $37.1$/$34.5$ \\ 
SwinUNet \cite{Swin-Unet}& $27.18$ & $30.67$ & $37.8$/$35.2$ \\ 
CA\textsuperscript{2.5}-Net \cite{CA2.5net}  & $24.27$ & $460.70$ & $73.8$/$70.2$  \\ 
\hline
Ours (w/o MLP \& w/o AS) & $34.33$ & $93.98$  &$76.1$/$74.3$ \\ 
Ours (w/o MLP) & $30.82$ & $123.60$ & $62.6$/$61.2$ \\ 
Ours (w/o AS)  & $21.33$ & $116.95$  & $53.1$/$51.2$ \\ 
Ours (full settings) & $17.82$ & $165.95$ & $51.5$/$50.8$\\ 
\bottomrule
\end{tabular}
\end{table}

\section{Experiments}
\subsubsection{Dataset.}
We evaluated the applicability of our approach across multiple modalities by conducting evaluations on microscopy and histology datasets.
(1) \textit{Fluorescence Microscopy Image Dataset}: 
This set combines three different data sources to simulate the heterogeneous nature of medical images~\cite{dataset_CA}.
It consists of 524 fluorescence images, each with a resolution of $512 \times 512$ pixels. 
(2) \textit{Histology Image Dataset}: 
This set is the combination of the open dataset MoNuSeg~\cite{MICCAI2018} and another private histology dataset~\cite{ClusterSeg} of $462$ images. 
We crop each image in the MoNuSeg dataset into four partially overlapping $512\times512$ images.
The private dataset contains $300$ images sized at $512\times512$ tessellated from 50 WSIs scanned at $20\times$, and meticulously labeled by five pathologists according to the labeling guidelines of the MoNuSeg~\cite{MICCAI2018}. 
For both datasets, we randomly split $80\%$ of the samples on the patient level as the training set and the remaining $20\%$ as the test set. 

\subsubsection{Implementations.}

All experiments are performed on one NVIDIA RTX $3090$ GPU with $24$ GB memory. 
We use Adam optimizer with an initial learning rate of $1\times10^{-4}$. 
We compare \texttt{TransNuSeg} with UNet~\cite{U-Net}, UNet++~\cite{Unet++}, TransUNet~\cite{Trans-Unet}, SwinUNet~\cite{Swin-Unet}, and CA\textsuperscript{2.5}-Net~\cite{CA2.5net}. 
We evaluate the results by using Dice Score (DSC), Intersection over Union (IoU), pixel-level accuracy (Acc), and F1-score(F1) as metrics, and ErCnt~\cite{ClusterSeg}.
To ensure statistical significance, we run all methods five times with different fixed seeds and report the results as mean $\pm$ standard deviation.

\begin{figure}[t!]
\centering
\includegraphics[width=0.95\textwidth]{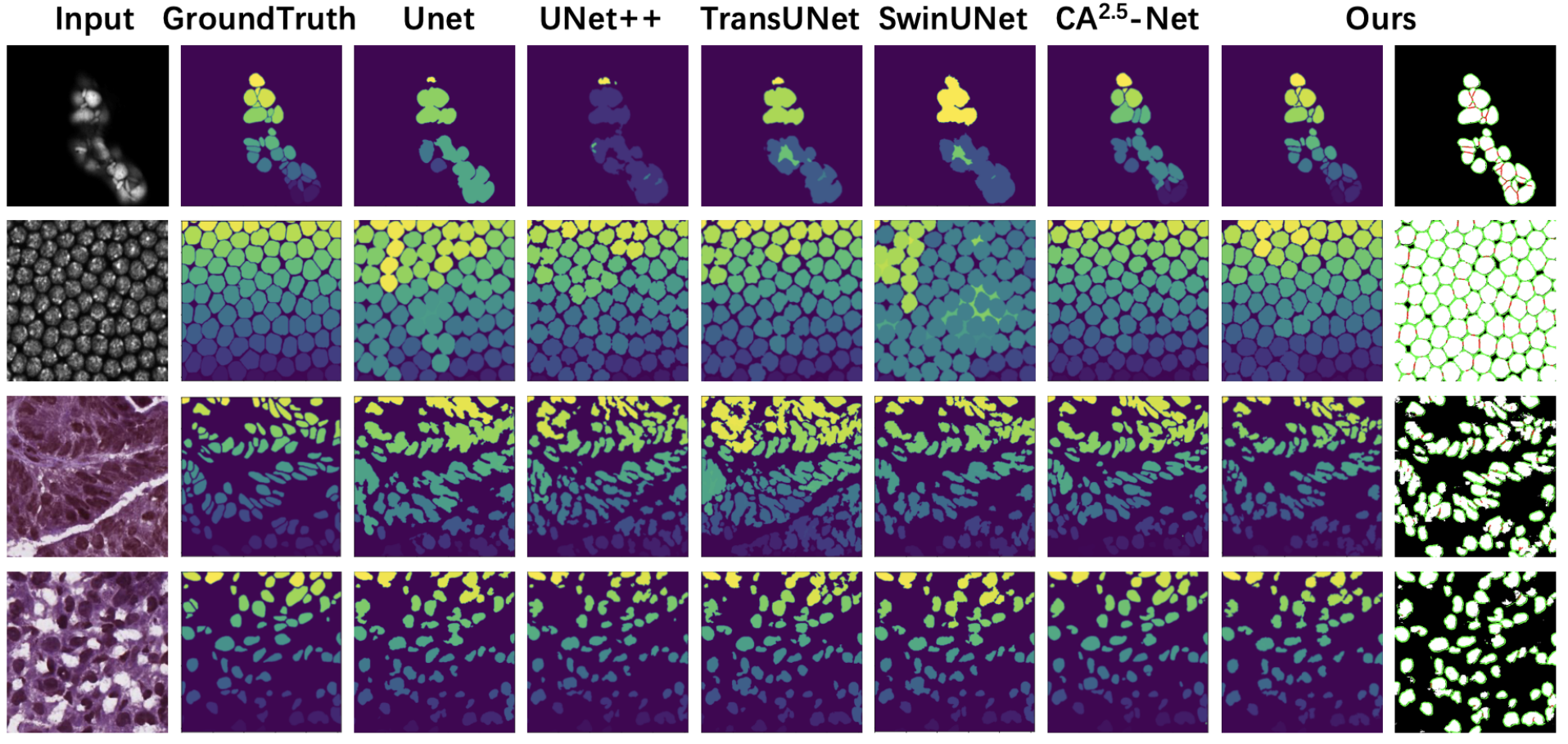}
\caption{
Exemplary samples and their segmentation results using different methods. 
\texttt{TransNuSeg} demonstrates superior segmentation performance compared to its counterparts, which can successfully distinguish severely clustered nuclei from normal edges.
}\label{fig:Result}
\end{figure}

\begin{figure}[b!]
\centering
\includegraphics[width=0.95\textwidth]{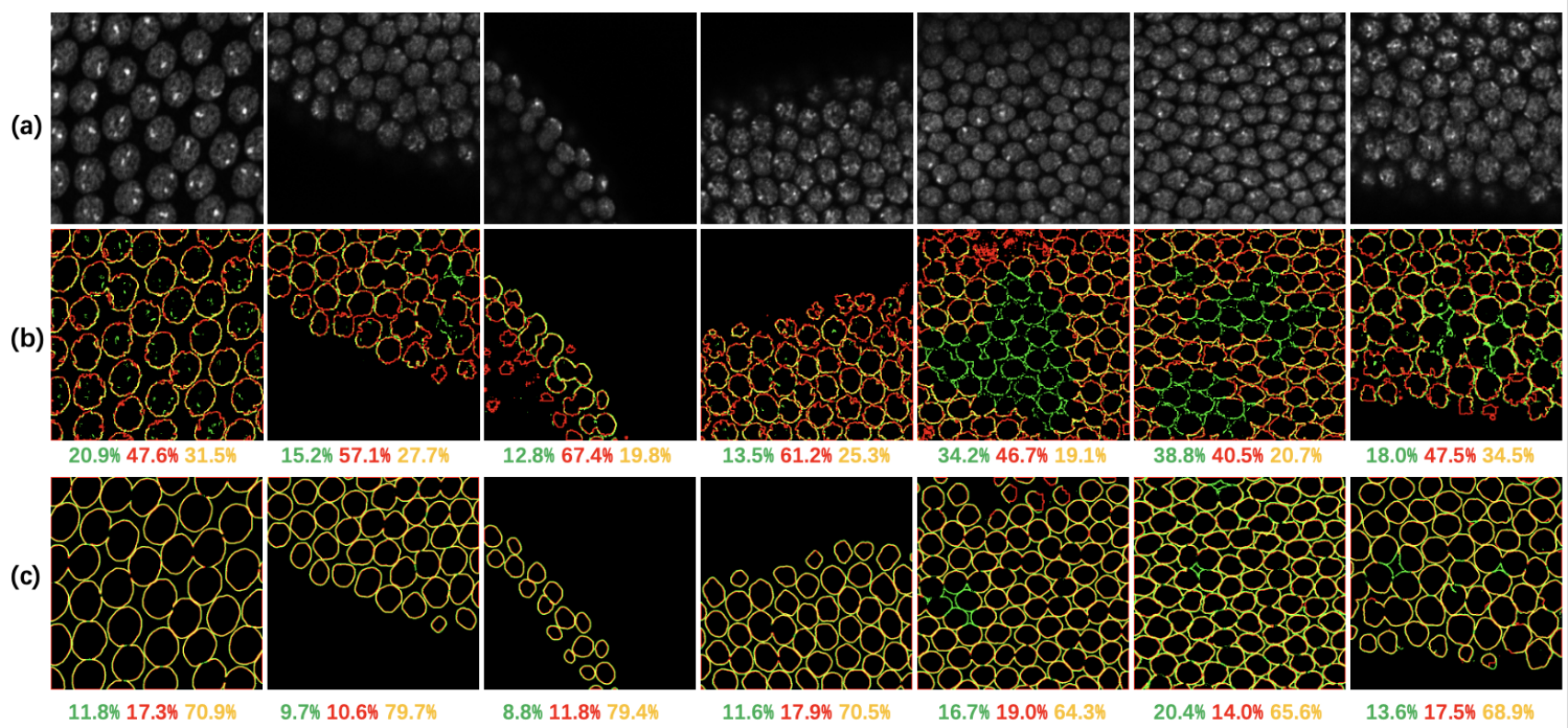}
\caption{
The impact of self distillation regularization on mismatch reduction across three decoders. 
(a) Raw input image. Segmentation results by \texttt{TransNuSeg} trained (b)  w/o self distillation, and (c) w/ self distillation.
The predicted normal edges from the normal edge decoder are shown in \textcolor{mygreen}{green}; while the edges generated from the nuclei decoder and processed with the Sobel operation are in \textcolor{myred}{red}. 
The \textcolor{myyellow}{yellow} color indicates the overlap between both. 
Accordingly, the numbers below images indicate the proportion of the pixels belonging to the three parts.
Compared to the results without self distillation, the outputs with self distillation exhibit reduced mismatches, resulting in improved segmentation performance.
}
\label{fig:CA_vis}
\end{figure}

\subsubsection{Results.}
Table \ref{tab:result} shows the quantitative comparisons for the nuclei segmentation.
The large margin between the SwinUNet and the other CNN-based or hybrid networks also confirms the superiority of the Transformer in fine-grained nuclei segmentation. 
More importantly, our method can outperform SwinUNet and the previous methods on both datasets.
For example, in the histology image dataset, \texttt{TransNuSeg} improves the dice score, F1 score, accuracy, and IoU by $2.08\%$, $3.41\%$, $1.25\%$, and $2.70\%$ respectively, over the second-best models. 
Similarly, in the fluorescence microscopy image dataset, our proposed model improves DSC by $0.96\%$, while also leading to $1.65\%$, $1.03\%$ and $1.91\%$ increment in F1 score, accuracy, and IoU to the second-best performance. 
For better visualization, representative samples and their segmentation results using different methods are demonstrated in Fig. \ref{fig:Result}. 
Furthermore, Table \ref{tab:cost} compares the model complexity in terms of the number of parameters, floating point operations per second (FLOPs), and the training computational cost, where our approach can significantly reduce around $28\%$ of the training time compared to the state-of-the-art CNN multi-task method CA\textsuperscript{2.5}-Net, while also boosting performance.

\begin{table}[t!]
\caption{
The ablation on each functional block, where `MLP', `AS', and `SD' represent the token MLP bottleneck, attention sharing scheme, and the self distillation. 
}
\centering
\label{tab:AblationStudy}
\begin{tabular}{ccc|cccc|cccc}
\hline
\multicolumn{1}{c}{\multirow{2}{*}{MLP}} & \multicolumn{1}{c}{\multirow{2}{*}{AS}} & \multicolumn{1}{c|}{\multirow{2}{*}{SD}} & \multicolumn{4}{c|}{Microscopy} & \multicolumn{4}{c}{Histology} \\ \cline{4-11} 
\multicolumn{1}{c}{}& \multicolumn{1}{c}{} & \multicolumn{1}{c|}{} 
& DSC (\%) & F1 (\%) & Acc (\%) & IoU (\%) & DSC (\%) & F1 (\%) & Acc (\%) & IoU (\%)\\ \hline
$\times$ & $\times$ & $\times$ &  $95.31$ & $94.05$ & $96.06$ & $90.05$ & $88.76$ & $78.20$ & $90.96$ & $64.48$  \\
$\bullet$ & $\times$ & $\times$ & $95.49$ & $94.48$ & $95.95$ & $89.97$ & $89.41$ & $77.94$ & $91.02$ & $65.17$ \\
$\times$ & $\bullet$ & $\times$ & $95.88$ & $93.51$ & $96.11$ & $90.55$ & $90.23$ & $80.46$ & $92.03$ & $67.84$\\
$\bullet$ & $\bullet$ & $\times$ & $96.95$ & $95.72$ & $96.92$ & $91.98$ & $90.27$ & $81.04$ & $92.01$ & $67.56$\\
$\times$ & $\times$ & $\bullet$ & $96.99$ & $95.74$ & $97.02$ & $92.22$ & $90.25$ & $80.81$ & $92.45$ & $68.14$\\
$\bullet$ & $\times$ & $\bullet$ & $96.58$ & $95.65$ & $97.03$ & $92.07$ & $90.17$ & $80.62$ & $92.35$ & $67.88$\\
$\times$  & $\bullet$ & $\bullet$ & $96.89$ & $95.78$ & $97.12$ & $92.08$ & $90.34$ & $80.88$ & $92.49$ & $68.05$ \\
$\bullet$ & $\bullet$ & $\bullet$  & \bm{$97.01$}  & \bm{$96.67$} & \bm{$97.11$} & \bm{$92.97$} & \bm{$90.81$} & \bm{$81.52$} & \bm{$92.77$} & \bm{$69.49$}\\ 
\hline
\end{tabular}
\end{table}

\subsubsection{Ablation.}
Our ablation study yields that token MLP bottleneck and attention sharing schemes can complementarily reduce the training cost while increasing efficiency, as shown in Table \ref{tab:cost} (the last 4 rows).
To further show the effectiveness of these schemes, as well as consistency self distillation, we conduct a  comprehensive ablation study on both datasets.
As described in Table~\ref{tab:AblationStudy}, each component proportionally contributes to the improvement to reach the overall performance boost.
Moreover, self distillation can enhance the intrinsic consistency between two branches, as visualized in Fig.~\ref{fig:CA_vis}.

\section{Conclusion}
In this paper, we make the first attempt at an efficient but effective multi-task Transformer framework for modality-agnostic nuclei segmentation. 
Specifically, our tri-decoder framework \texttt{TransNuSeg} leverages an innovative self distillation regularization to impose consistency between the different branches.
Experimental results on two datasets demonstrate the excellence of our \texttt{TransNuSeg} against state-of-the-art counterparts for potential real-world clinical deployment. 
Additionally, our work opens a new architecture to perform nuclei segmentation tasks with Swin Transformer, where further investigations can be performed to explore the generalizability to the top of our methods with different modalities.

\bibliographystyle{plain}
\bibliography{0miccai/6_ref}

\end{document}